\documentclass[english,aps,twocolumn,prX,showpacs,superscriptaddress]{revtex4-1}

\usepackage{graphicx}
\usepackage{dcolumn}
\usepackage{bm}
\usepackage{babel}
\usepackage[T1]{fontenc}
\usepackage[latin9]{inputenc}
\usepackage{amsmath}
\usepackage{amssymb}
\usepackage{amsfonts}
\usepackage{booktabs}
\usepackage{microtype}
\usepackage[version=4]{mhchem}
\DeclareUnicodeCharacter{2212}{-} 
\usepackage{hyperref}
\usepackage{siunitx}
\setcitestyle{numbers,square}
\usepackage{color}
\hypersetup{
	colorlinks = true,
	pdftitle = {},
	pdfauthor = {},
	pdfkeywords = {},
	linkcolor = blue,
	citecolor = blue,
	filecolor = black,
	urlcolor = magenta,
}

\begin{document}
\title{Strain-tunable interface electrostatics in Janus MoSSe/silk vdW heterostructure for triboelectric nanogeneration}

\author{Deobrat Singh}
\email{deosing@kth.se}
\affiliation{Department of Materials Science and Engineering, KTH The Royal Institute of Technology, Stockholm SE-100 44, Sweden}
\author{Raquel Liz\'arraga}
\email{raqli@kth.se}
\affiliation{Department of Materials Science and Engineering, KTH The Royal Institute of Technology, Stockholm SE-100 44, Sweden}
\affiliation{Wallenberg Initiative Materials Science for Sustainability, Department of Materials Science and Engineering, KTH The Royal Institute of Technology, Stockholm SE-100 44, Sweden}

\date{\today}

\begin{abstract}
Understanding and engineering interfacial electrostatics in hybrid two-dimensional (2D) and biomolecular material systems is essential for advancing high-performance triboelectric nanogenerators (TENGs). In this work, we systematically investigate the strain-dependent electronic structure and triboelectric response of Janus MoSSe, silk fibroin, and their van der Waals (vdW) heterostructure using first-principles calculations. Tensile strain induces a pronounced band-gap reduction in the MoSSe/silk interface, exceeding that of the isolated constituents and indicating enhanced interlayer electronic coupling. The vdW heterostructure exhibits a significant work-function shift and a substantially larger dipole moment compared to MoSSe and silk alone, revealing strong interfacial charge redistribution driven by Fermi-level alignment and asymmetric polarization. This enhanced polarization directly amplifies the triboelectric surface charge density, producing values more than double those of pristine MoSSe and several orders of magnitude higher than silk. Consequently, the open-circuit voltage and overall triboelectric output are markedly improved across all strain levels. These results demonstrate that synergistic interfacial polarization and strain engineering can effectively elevate charge separation, storage, and transfer efficiencies, establishing the MoSSe/silk vdW heterostructure as a promising material for next-generation high-efficiency TENGs.
\end{abstract}

\maketitle


\section{Introduction}\label{sec:intro}
The rapid expansion of portable, wearable, and self-powered electronic systems has significantly increased the demand for efficient mechanical energy harvesting technologies. Among various approaches, TENGs and piezoelectric nanogenerators (PENGs) have emerged as promising candidates for converting ambient mechanical energy into electrical output \cite{wang2013triboelectric,wang2014piezoelectric}. While PENGs rely on strain-induced polarization in non-centrosymmetric materials, TENGs operate through contact electrification and electrostatic induction between materials with dissimilar electron affinities. Recent studies suggest that integrating triboelectric and piezoelectric mechanisms within a single architecture can significantly enhance charge density, broaden frequency response, and improve overall power output \cite{zhu2016hybrid,yang2015hybrid}. Such hybrid nanogenerators are particularly attractive for flexible, wearable, and bio-integrated applications\cite{wang2017triboelectric}.

Despite substantial experimental progress, the microscopic origin of coupled triboelectric and piezoelectric effects remains incompletely understood. Triboelectric charge generation fundamentally originates from atomic-scale charge transfer governed by work function differences and interfacial electronic hybridization \cite{wang2019origin}. Similarly, piezoelectricity in low-dimensional materials is strongly dependent on symmetry breaking and strain-induced modulation of electronic states. First-principles density functional theory (DFT) calculations have therefore become indispensable tools for predicting polarization behavior, interfacial charge redistribution, and strain-dependent electronic properties prior to experimental realization \cite{giannozzi2009quantum,kresse1996efficient}. Computational materials design enables systematic screening of layered materials and heterostructures, reducing the need for extensive experimental trial-and-error approaches.

2D materials offer unique advantages for hybrid nanogenerators due to their mechanical flexibility, high surface-to-volume ratio, and tunable electronic properties. Symmetric monolayer transition metal dichalcogenides (TMDs) such as \ce{MoS2}, \ce{WS2}, \ce{MoSe2} and monochalcogenides possess mirror symmetry with identical chalcogen atoms on both sides of the transition metal plane, resulting in the absence of intrinsic out-of-plane dipole moments \cite{mak2010atomically,splendiani2010emerging, fei2015giant,wu2014piezoelectricity,dong2017piezoelectricity,zhang2018strain}. Although monolayer \ce{MoS2} lacks inversion symmetry and exhibits in-plane piezoelectricity, its overall vertical symmetry suppresses spontaneous out-of-plane polarization \cite{wu2014piezoelectricity,dong2017piezoelectricity,dong2017piezoelectricity}. This structural symmetry limits their ability to generate built-in electric fields compared to Janus counterparts. Nevertheless, symmetric TMD monolayers have been extensively studied both computationally and experimentally for applications in nanoelectronics, optoelectronics, and electromechanical devices due to their direct bandgap, strong spin-orbit coupling, and mechanical flexibility \cite{radisavljevic2011single,wang2012electronics}. The transition from symmetric to asymmetric (Janus) structures introduces vertical polarity and enhanced piezoelectric response, offering improved performance for energy harvesting applications \cite{dong2017piezoelectricity}. In particular, Janus TMDs such as MoSSe exhibit intrinsic out-of-plane asymmetry arising from different chalcogen atoms on opposite sides of the monolayer \cite{zhang2017janus,lu2017janus}. This broken mirror symmetry generates a built-in dipole moment and enhances piezoelectric polarization, making Janus MoSSe an excellent candidate for piezoelectric nanogenerators \cite{li2018intrinsic,lu2017janus}. Experimental synthesis and characterization have confirmed the structural stability and polarity of MoSSe monolayers, further validating their practical relevance \cite{zhang2017janus}.

Beyond inorganic 2D materials, biomaterials such as silk in its $\beta$-sheet ($\beta$-form) configuration have attracted attention for triboelectric applications due to their mechanical robustness, biocompatibility, and intrinsic dipolar hydrogen-bonded network \cite{kim2018silk}. The integration of 2D materials with polymers has demonstrated enhanced triboelectric and piezoelectric responses in flexible nanogenerators \cite{wu2020moS2}. The vdW heterostructures provide a versatile platform for combining such materials without stringent lattice matching constraints \cite{zhang2019vdw,yu2018vdw}. At the quantum level, vdW heterostructures provide a powerful platform for nanogenerator design because they allow tunable interfacial charge transfer and built-in electric fields while preserving the intrinsic electronic properties of individual layers \cite{geim2013vdw,novoselov2016vdw}. The weak interlayer coupling minimizes lattice strain and defect formation, yet enables efficient modulation of band alignment, electrostatic potential, and polarization under mechanical deformation \cite{yu2018vdw,zhang2019vdw}. Such controllable interfacial electrostatics are crucial for enhancing coupled triboelectric and piezoelectric responses, where charge redistribution and polarization screening occur at the atomic scale \cite{wang2019origin,baroni2001dfpt}. Furthermore, the absence of dangling bonds at vdW interfaces reduces trap-assisted recombination and improves charge stability, which is beneficial for sustained energy conversion performance. However, experimental realization still faces challenges, including precise control of interfacial cleanliness, atomic registry, and strain transfer efficiency. Direct measurement of atomic-scale triboelectric charge redistribution and interfacial polarization remains limited by current characterization techniques, underscoring the importance of first-principles investigations.

Motivated by these considerations, we propose a vdW heterostructure composed of Janus MoSSe and $\beta$-form of silk as a hybrid tribo-piezoelectric nanogenerator platform. The intrinsic piezoelectric response of MoSSe is expected to synergistically couple with interfacial triboelectric charge transfer at the MoSSe/silk interface. Using first-principles DFT calculations, we systematically investigate the structural stability, electronic properties, strain-dependent polarization, and interfacial charge redistribution mechanisms of the proposed heterostructure. Our study provides atomic-scale insights into hybrid nanogenerator design and offers computational guidelines for next-generation flexible energy harvesting systems.


\section{Computational approach}
\label{method}
All simulations in this study were performed using density functional theory (DFT), a quantum mechanical method widely used to investigate the electronic properties of materials. The theoretical framework follows the Hohenberg-Kohn formulation, which establishes that the ground-state properties of a many-electron system can be uniquely determined from its electron density \cite{hohenberg1964inhomogeneous}. The interaction between valence electrons and ionic cores was described using the projector augmented wave (PAW) method, as implemented in the Vienna \textit{Ab-Initio} Simulation Package (VASP) \cite{kresse1996efficient}. The exchange-correlation energy was treated using the generalized gradient approximation (GGA) with the Perdew-Burke-Ernzerhof (PBE) functional. Since the investigated systems involve heterostructures where weak interlayer interactions play an important role, long-range van der Waals (vdW) dispersion corrections were included to properly describe the interaction between layers \cite{frostenson2021hard,hyldgaard2020screening}.

The electronic wave functions were expanded in a plane-wave basis set with a kinetic energy cutoff of 500 eV, which was selected after performing convergence tests to ensure reliable results. Structural optimizations were carried out until the total energy difference between successive ionic steps became smaller than $10^{-8}$ eV. At the same time, the residual Hellmann-Feynman forces acting on each atom were reduced below $10^{-3}$ eV/\AA, ensuring that the optimized structures correspond to stable equilibrium configurations. For Brillouin zone sampling, a $\Gamma$-centered $18\times18\times1$ and $8\times10\times8$ $k$-point grid was employed, which is appropriate for describing the electronic properties of the two-dimensional Janus MoSSe monolayer and bulk crystalline silk polymer unit cell considered in this work. The electronic density of states (DOS) was calculated using the Gaussian smearing technique with a broadening width of 0.12 eV in order to obtain smooth and well resolved spectra. To avoid artificial interactions between periodically repeated images, a vacuum layer of approximately 20 \AA\ was introduced along the direction transverse to the surface for both the isolated unit cells and the vdW heterostructure model. Charge redistribution among the constituent atoms was evaluated using the Bader charge analysis method \cite{henkelman2006fast}, which allows the electronic charge associated with each atom to be quantified. Furthermore, the nature of chemical bonding within the system was examined through crystal orbital Hamilton population (COHP) analysis using the LOBSTER package \cite{maintz2016lobster}. This method provides quantitative insight into bonding and antibonding interactions between specific atomic pairs, enabling a deeper understanding of the bonding characteristics in the studied structures. 

\textit{Ab-initio} molecular dynamics (AIMD) simulations of the Janus MoSSe/silk vdW heterostructure were performed in the NVT ensemble at 300 K using a Nos\'e-Hoover chain thermostat. Ionic positions were propagated with a time step of 1 fs over 5000 steps without structural relaxation, while electronic self-consistency was converged to $10^{-8}$ eV per step. The total energy remained stable throughout the trajectory, confirming proper equilibration and the dynamical stability of the vdW heterostructure.

\section{Results and Discussion}
\subsection{Structural Information and Stability} \label{structural-prop}
\begin{figure}[htp!]
	    \centering
	    \includegraphics[width=1.0\linewidth]{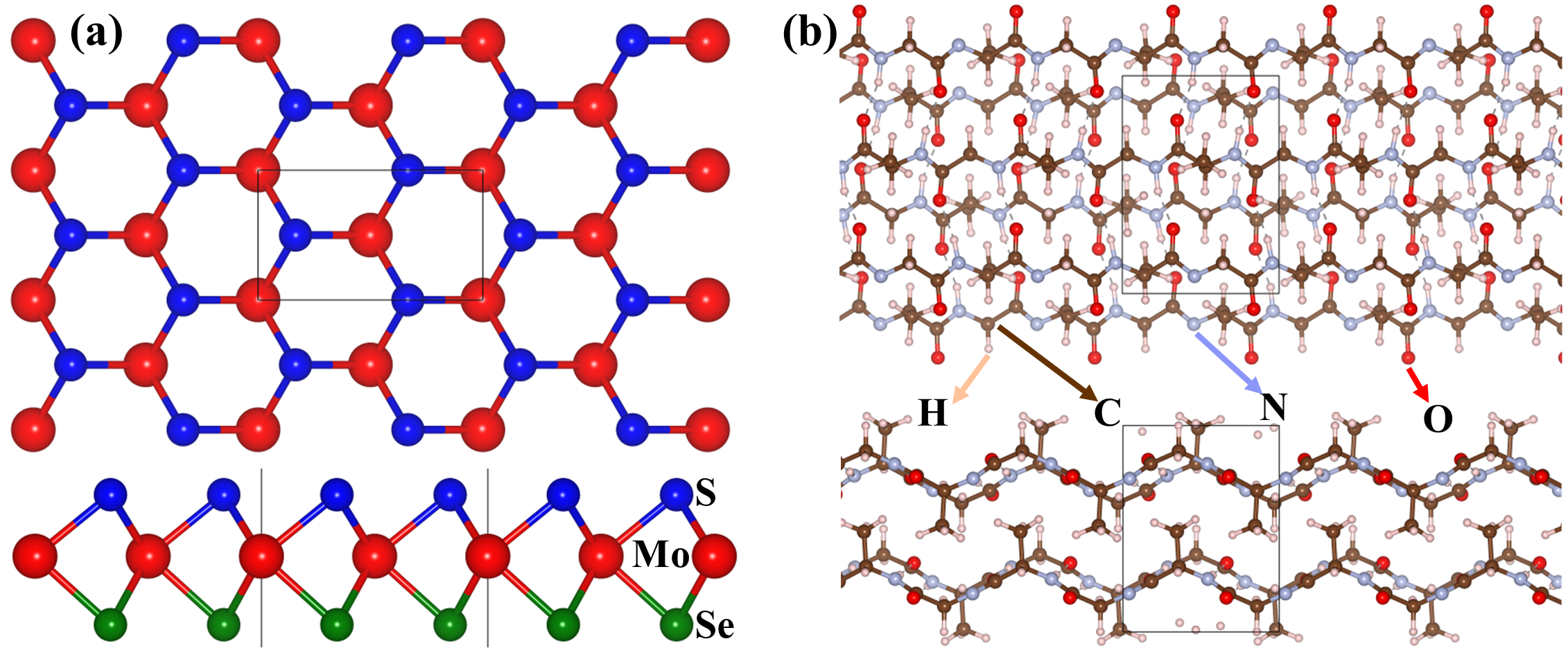}
	    \caption{Optimized atomic structures of the constituent materials used in this study. (a) Top view and side view of the Janus 2D MoSSe monolayer, illustrating the asymmetric arrangement of S and Se atoms on opposite sides of the Mo layer. (b) Top view and side view of the crystalline silk polymer structure (silk-$\beta$ phase). Different colored spheres represent the respective atomic species in the structures.}
	    \label{f1}
\end{figure}

The atomic configuration of the Janus MoSSe monolayer was constructed using a hexagonal lattice in which a Mo atom is positioned between two different chalcogen layers composed of S and Se atoms. This asymmetric arrangement breaks the out-of-plane mirror symmetry that is typically present in conventional transition metal dichalcogenides, giving rise to the distinctive properties of Janus structures \cite{lu2017janus,zhang2017janus}. After structural optimization, the lattice constants were obtained as $a=b=3.25$~\AA (see Figure \ref{f1}a). These values are consistent with previously reported theoretical and experimental studies of Janus MoSSe systems \cite{lu2017janus,li2018janus}.

In the optimized structure, the Mo atom forms covalent bonds with the surrounding chalcogen atoms on both sides of the layer. The calculated bond lengths are approximately 2.42~\AA\ for Mo-S and 2.54~\AA\ for Mo-Se. The difference in these bond distances originates from the different atomic radii and bonding characteristics of sulfur and selenium atoms. The resulting structural asymmetry generates an intrinsic dipole moment along the out-of-plane direction, which is a characteristic feature of Janus two-dimensional materials \cite{zhang2017janus,dong2018intrinsic}. To eliminate artificial interactions between periodically repeated layers, a vacuum region of about 15~\AA\ was introduced along the perpendicular direction. Electronic structure calculations further reveal that the MoSSe monolayer behaves as a semiconductor with an estimated band gap of approximately 1.59~eV, which agrees well with earlier theoretical reports for Janus transition metal dichalcogenides \cite{li2018janus,dong2018intrinsic}.

For the silk component, the crystal structure corresponds to the Silk-$\beta$ phase, which is associated with the $\beta$-sheet configuration commonly observed in natural silk fibers \cite{marsh1955structure,porter2018silk}. The optimized lattice parameters were determined to be $a=9.36$~\AA, $b=6.74$~\AA, and $c=8.91$~\AA (see Figure \ref{f1}b. The unit cell contains N, H, C and O atoms arranged in a periodic molecular network that reflects the polypeptide backbone of the silk protein chains. The stability of this structure mainly arises from strong covalent bonding within the molecular chains and hydrogen bonding interactions between adjacent chains, which together provide the characteristic mechanical strength of silk materials \cite{porter2018silk,asakura2020silk}. The electronic structure calculations indicate that silk behaves as a wide band-gap insulating material with a band gap of 4.44~eV. The combination of a semiconducting Janus MoSSe monolayer and an insulating silk structure forms a promising platform for constructing vdW heterostructures. The difference in electronic properties between these two materials can facilitate charge redistribution and interface polarization, which are important factors for designing functional nanoscale electronic and energy-harvesting devices.

\subsection{Chemical Bonding and Structural Stability}\label{chemical-bond}

\begin{figure}[htp!]
	    \centering
	    \includegraphics[width=1.0\linewidth]{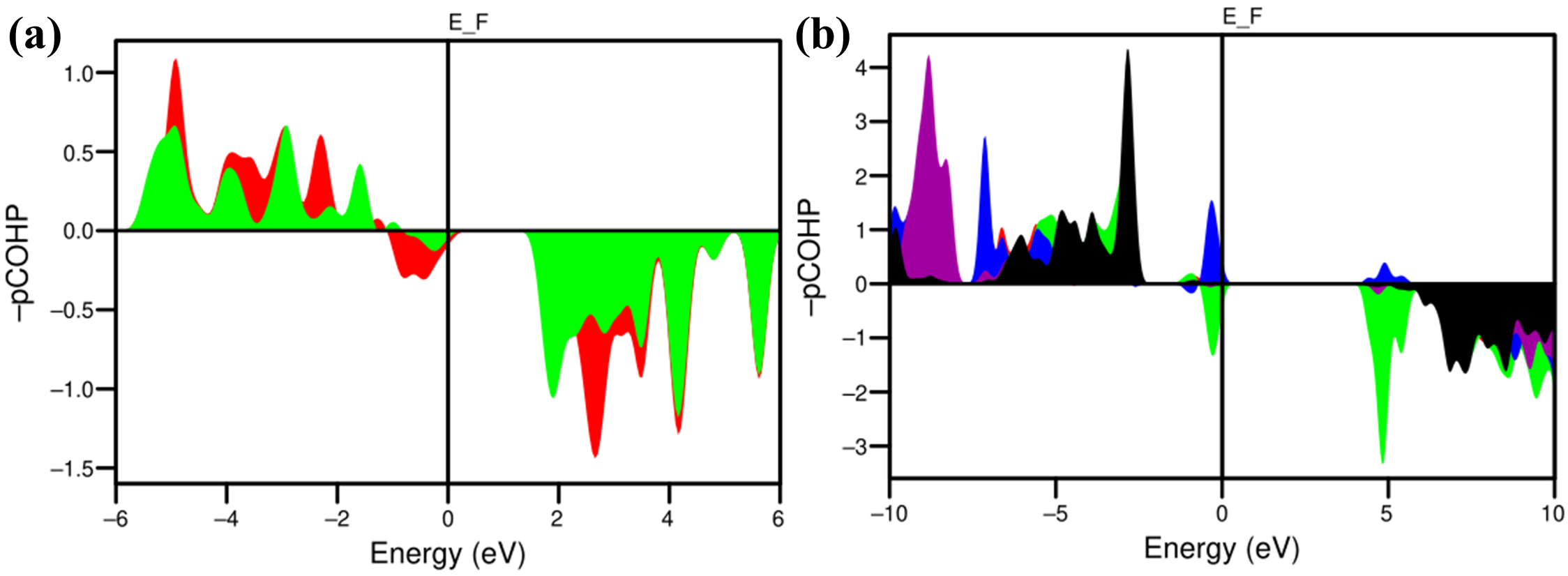}
	    \caption{Crystal orbital Hamilton population (COHP) analysis of chemical bonding. (a) COHP curves for the Janus MoSSe monolayer showing Mo-S (red) and Mo-Se (green) bonding interactions. (b) COHP curves for the Silk-$\beta$ polymer structure illustrating the bonding contributions from N-C (red), C-O (green), C-C (blue), N-H (violet), and H-C (black) atomic pairs. The vertical line denotes the Fermi level ($E_F$). Negative values correspond to bonding states, while positive values indicate antibonding interactions.}
	    \label{cohp}
    \end{figure}

To gain deeper insight into the bonding nature and structural stability of the studied systems, the chemical interactions between atomic pairs were analyzed using the crystal orbital Hamilton population (COHP) method. This approach provides an energy-resolved description of bonding and antibonding interactions between atoms and is widely used to understand chemical bonding in solids. In COHP analysis, (-)ve values correspond to bonding states, whereas (+)ve values indicate antibonding interactions. The integrated COHP (ICOHP) values quantify the overall bond strength, where more negative values indicate stronger bonding interactions \cite{dronskowski1993cohp,maintz2016lobster}. Figure~\ref{cohp}a shows the COHP curves for the Janus MoSSe monolayer. In this structure, the Mo atom is coordinated with S and Se atoms located on opposite sides of the layer, which breaks the out-of-plane symmetry characteristic of conventional transition metal dichalcogenides. The red and green shaded regions represent the Mo-S and Mo-Se bonding interactions, respectively. The optimized bond lengths are found to be 2.42~\AA\ for Mo-S and 2.54~\AA\ for Mo-Se. The corresponding integrated COHP values are $-3.08$ eV and $-2.84$ eV. These negative ICOHP values indicate strong covalent bonding between the Mo atom and the surrounding chalcogen atoms. The slightly larger magnitude of the Mo-S ICOHP value suggests that the Mo-S bond is marginally stronger than the Mo-Se bond, which is consistent with its shorter bond length and stronger orbital overlap \cite{lu2017janus,li2018janus}.

The COHP results for the crystalline Silk-$\beta$ polymer structure are presented in Fig.~\ref{cohp}b. Several atomic interactions contribute to the stability of the polymeric network, including N-C, C-O, C-C, N-H, and H-C bonding pairs. These interactions are represented by red, green, blue, violet, and black shaded regions, respectively. The calculated bond lengths are approximately 1.34~\AA\ for N-C, 1.24~\AA\ for C-O, 1.53~\AA\ for C-C, 1.02~\AA\ for N-H, and 1.09~\AA\ for H-C. The corresponding ICOHP values are $-13.13$, $-16.59$, $-9.27$, $-8.41$, and $-7.21$ eV, respectively. Among these interactions, the C-O bond exhibits the most negative ICOHP value, indicating the strongest bonding interaction within the silk polymer structure. Strong covalent bonding contributions from N-C and C-C interactions further stabilize the polypeptide backbone of the silk chains. Meanwhile, the N-H and H-C interactions contribute to additional stabilization through covalent bonding and intermolecular hydrogen-bond networks, which are known to play an essential role in maintaining the mechanical robustness and stability of the Silk-$\beta$ crystalline phase \cite{marsh1955structure,porter2018silk}. It means that, the COHP analysis demonstrates that both the Janus MoSSe monolayer and the Silk-$\beta$ polymer possess strong bonding interactions that ensure structural stability. The covalent Mo-S and Mo-Se bonds maintain the integrity of the Janus 2D lattice, while the strong covalent and hydrogen-bonding interactions in the silk polymer stabilize the organic framework. These stable bonding characteristics are essential for the formation and reliability of the proposed vdW heterostructure.

\subsection{Mechanical Stability and Elastic Properties}
Mechanical stability is an important requirement for materials intended for nano-energy harvesting applications, since nanogenerators rely on repeated mechanical deformation during operation. To evaluate the mechanical robustness of the constituent materials, the elastic constants were calculated using the stress-strain method as implemented in the VASPKIT post-processing package \cite{wang2021vaspkit}. The obtained elastic parameters were further analyzed to determine the stability conditions and mechanical behavior of the structures. For the Janus MoSSe monolayer, the calculated elastic tensor corresponds to a 2D hexagonal crystal system with two independent elastic constants, namely $C_{11}$ and $C_{12}$. The obtained elastic constants are $C_{11}=148.67$ N/m and $C_{12}=43.94$ N/m, while the shear elastic constant $C_{66}$ is derived from the relation $C_{66}=(C_{11}-C_{12})/2$, yielding a value of 52.36 N/m. According to the mechanical stability criteria for 2D hexagonal systems, the conditions $C_{11} > 0$ and $C_{11} > |C_{12}|$ must be satisfied for structural stability \cite{wang2022elastic}. The calculated values clearly satisfy these requirements, confirming that the Janus MoSSe monolayer is mechanically stable. The in-plane mechanical properties further reveal that the Young's modulus of the monolayer is approximately 135.68 N/m, while the shear modulus is 52.36 N/m and the Poisson's ratio is 0.296. Interestingly, the minimum and maximum values of these quantities are identical, indicating that the monolayer exhibits nearly isotropic mechanical behavior within the plane of the material. Such isotropic elasticity is advantageous for nanogenerator applications because it allows the material to respond uniformly to mechanical strain applied along different in-plane directions.

For the crystalline Silk-$\beta$ polymer structure, the elastic tensor corresponds to a monoclinic crystal system containing thirteen independent elastic constants. The calculated stiffness matrix shows relatively large elastic coefficients, with $C_{11}=585.28$ GPa, $C_{22}=298.39$ GPa and $C_{33}=157.12$ GPa, indicating strong resistance to deformation along the principal crystallographic directions. All eigenvalues of the stiffness matrix are positive, which satisfies the Born mechanical stability criteria for monoclinic crystals \cite{mouhat2014elastic}. This confirms that the optimized silk structure is mechanically stable.

The averaged polycrystalline elastic properties obtained from the Voigt-Reuss-Hill approximation provide additional insight into the macroscopic mechanical behavior. The calculated bulk modulus, Young's modulus, and shear modulus are 133.61 GPa, 254.63 GPa, and 107.68 GPa, respectively. The calculated Pugh ratio ($B/G$) is approximately 1.24, which is lower than the conventional ductile-brittle transition value of 1.75 \cite{pugh1954relations}. This result suggests that the bonding interactions in the silk structure are predominantly directional and covalent in nature, contributing to the structural rigidity and mechanical stability required for maintaining the integrity of the material under mechanical deformation. The elastic anisotropy analysis shows moderate anisotropic behavior for the silk polymer, with a universal elastic anisotropy index of 1.38. Such anisotropy originates from the complex molecular arrangement and hydrogen-bond network present in the Silk-$\beta$ crystalline phase. Despite this anisotropy, the material exhibits relatively high elastic moduli, reflecting its well-known mechanical strength and structural rigidity. Finally, the calculated elastic constants and stability criteria confirm that both the Janus MoSSe monolayer and the Silk-$\beta$ polymer possess sufficient mechanical stability. The combination of mechanically robust inorganic and organic components is beneficial for the formation of a stable vdW heterostructure. Moreover, the ability of these materials to sustain mechanical deformation without structural failure is particularly advantageous for nanogenerator applications, where repeated strain cycles are required to convert mechanical energy into electrical energy.

    \begin{figure}[htp!]
	    \centering
	    \includegraphics[width=0.99\linewidth]{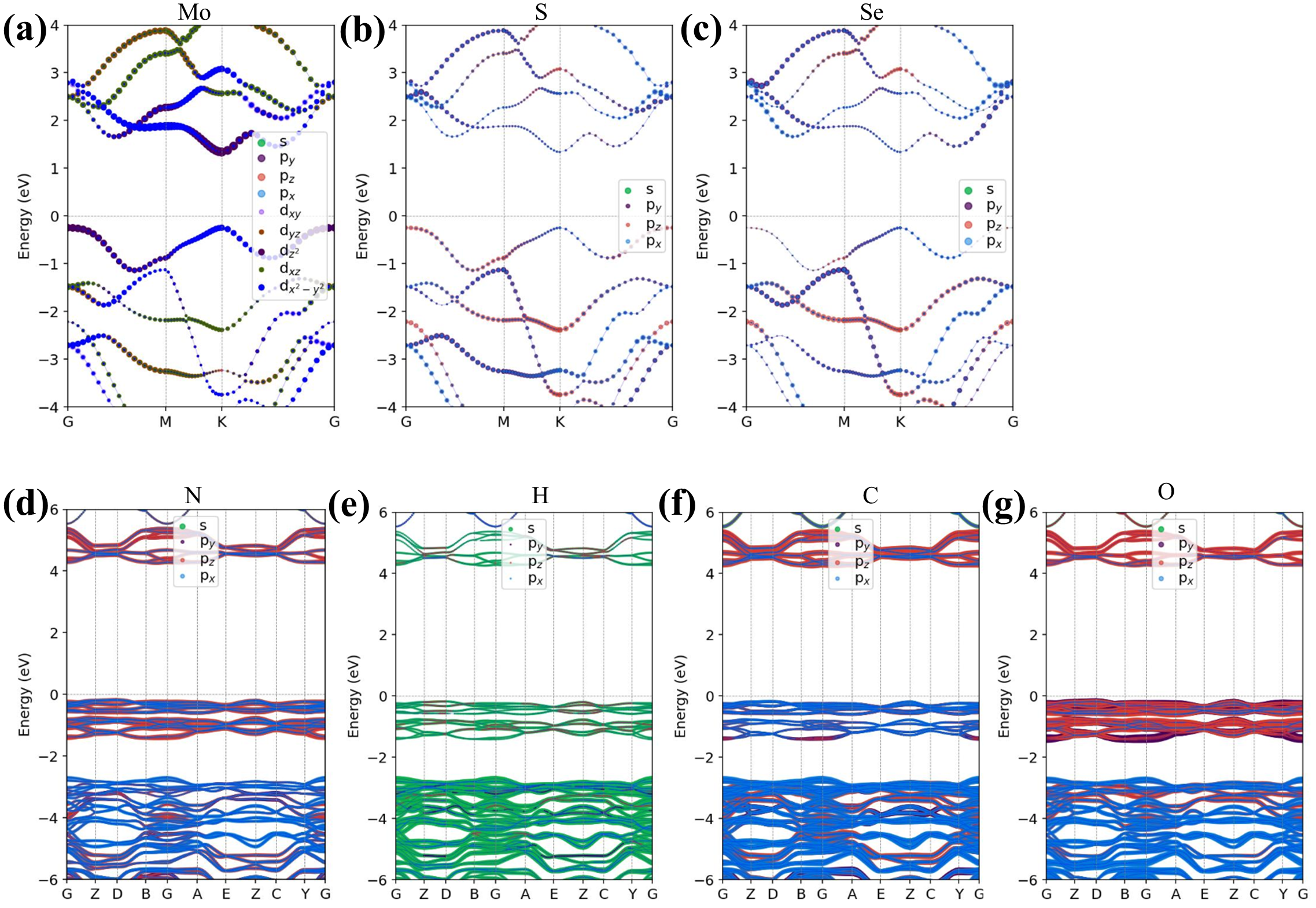}
	    \caption{Orbital-resolved electronic band structures for pristine Janus 2D MoSSe monolayer (a-c) and pure silk type-$\beta$ polymer structure (d-g).}
	    \label{f3}
    \end{figure}

Figure \ref{f3} and Figure S1 (see in the Supporting Information (SI)) shows the electronic band structures of pristine Janus 2D MoSSe monolayer and pristine silk type-$\beta$ polymer structure. Orbital-resolved electronic band structures of the pure Janus 2D monolayer containing Mo-S-Se atoms as presented in Figure \ref{f3}(a-c). The Figure \ref{f3}(a) shows the contributions from Mo-orbitals, highlighting the dominant d-orbital character near the conduction band minimum (CBM) and valence band maximum (VBM). It was seen that the Figure \ref{f3}(b and c) illustrate the S and Se contributions, respectively, where the p-orbitals primarily contribute to the valence bands. The band structures reveal the semiconducting nature of the monolayer, with the CBM and VBM located at the K-point, indicating a direct band gap. The projected density of states also confirm that Mo-d orbitals have main contribution near the Fermi-level as shown in Figure S2(a) in SI. Further, Figure \ref{f3}(d-g) shows the orbital-projected band structures of the pristine silk type-$\beta$ polymer structure containing N, H, C, and O atoms. Figure \ref{f3}(d) shows the nitrogen contributions, highlighting the localized states in the valence band region. Figure \ref{f3}(e) depicts hydrogen contributions, which are mostly minor but influence the electronic dispersion in the upper valence and lower conduction regions. Figure \ref{f3}(f and g) show the carbon and oxygen contributions, respectively, with p-orbitals dominating the valence band and key conduction states. The polymer exhibits a wider band gap compared to the 2D monolayer, with relatively flat bands reflecting the quasi-1D electronic nature along the polymer backbone. The density of states near the Fermi level is dominated by C and O p-orbitals indicating their critical role in the electronic structure and potential charge transport pathways as presented in Figure S2(b) (see in SI). This comparison emphasizes differences in orbital contributions, band dispersion, and dimensionality effects, which are crucial for understanding their triboelectric charge generation, piezoelectric response, and mechanical-to-electrical energy conversion efficiency in nanogenerator applications.

\subsection{Structural and electronic properties of Janus MoSSe/silk vdW heterostructure}
\begin{figure}[htp!]
\centering
\includegraphics[width=0.99\linewidth]{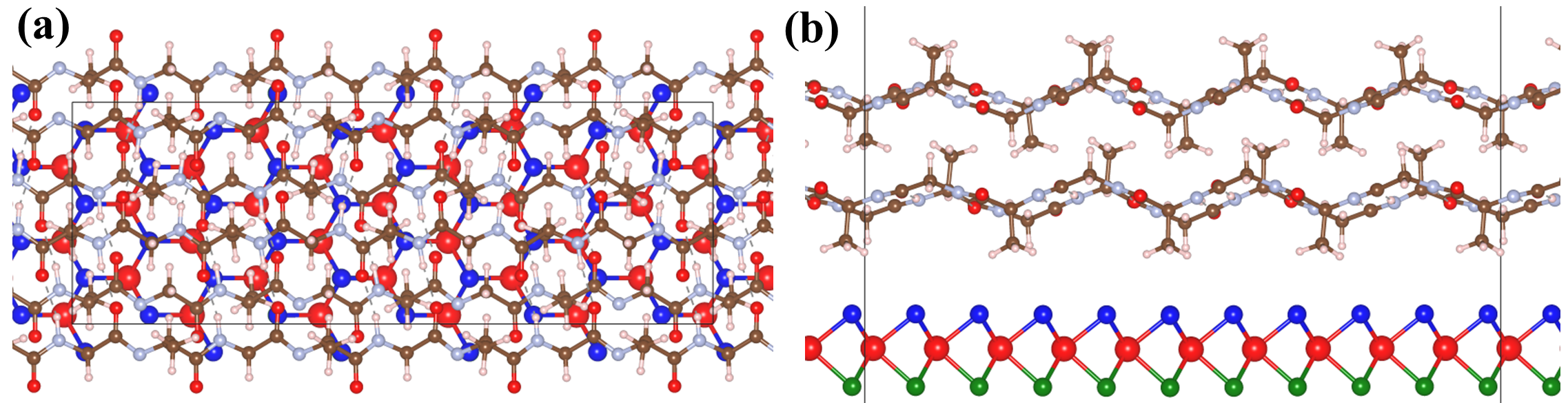}
\caption{Optimized atomic configuration (a) top view and (b) side view of the Janus MoSSe/silk vdW heterostructure.}
\label{vdw-str}
\end{figure}

Prior to interface formation, the isolated Janus MoSSe monolayer exhibits intrinsic Mo-S and Mo-Se bond lengths of approximately 2.42~\AA{} and 2.54~\AA{}, respectively, while the isolated silk polymer chain shows typical N-C, C-C, C-O, N-H, and C-H bonds close to 1.34-1.45~\AA{}, 1.53~\AA{}, 1.24~\AA{}, 1.02~\AA{}, and 1.09~\AA{}. After forming the heterostructure and performing full relaxation, both layers largely preserve their intrinsic bonding characteristics, confirming that the interaction is dominated by vdW forces. The interface stabilizes at a vertical separation of 2.55~\AA{}, with Mo-S and Mo-Se bonds optimized to 2.41~\AA{} and 2.52~\AA{}, respectively. The silk polymer exhibits N-C (1.36~\AA{} and 1.46~\AA{}), C-O (1.25~\AA{}), C-C (1.52-1.55~\AA{}), N-H (1.03~\AA{}), and C-H (1.11~\AA{}) bond lengths, all closely matching their isolated values. The heterostructure supercell is defined by lattice parameters of $a = 9.755$~\AA{}, $b = 28.16$~\AA{}, and $c = 36$~\AA{}, with a total thickness of 15.67~\AA{}.We have taken 20 \AA{} vacuum along the transverse direction to prevent the physical interaction between the periodic images. Lattice mismatches of 4\% along $a$ and 4.22\% along $b$ are accommodated without significant distortion, resulting in a stable and coherent vdW interface (see Figure \ref{vdw-str}), providing a reliable structural foundation for subsequent analysis of charge transfer, interface electrostatics, and triboelectric behavior. Figure S3 (see in SI) shows the time evolution of the total energy for the Janus MoSSe/silk vdW heterostructure at 300 K, obtained using \textit{ab-initio} molecular dynamics within the NVT ensemble (Nos\'e-Hoover chain). The small fluctuations without any noticeable energy drift indicate proper thermal equilibration and confirm the dynamical stability of the vdW heterostructure.

\begin{figure}[htp!]
	    \centering
	    \includegraphics[width=0.99\linewidth]{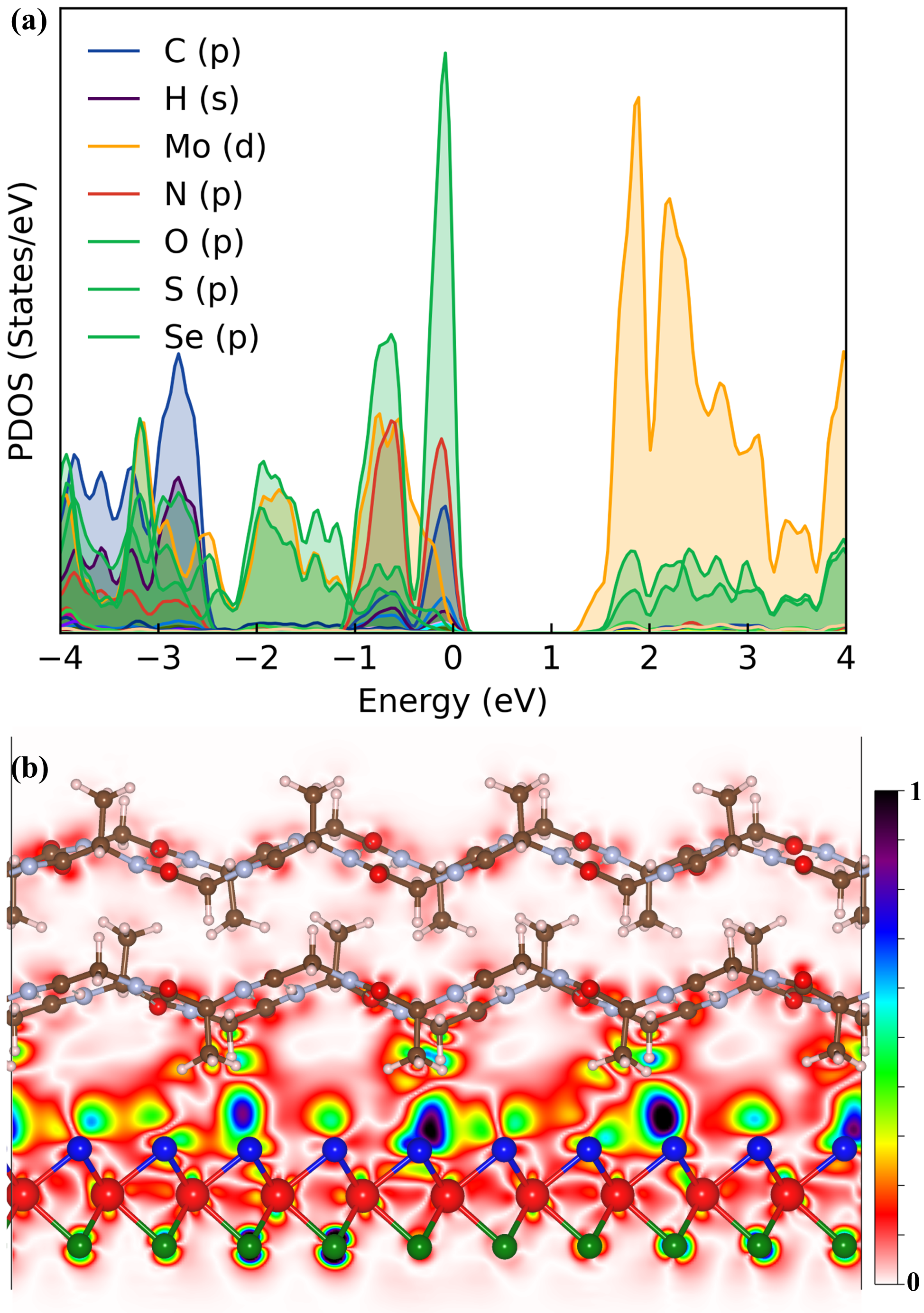}
	    \caption{(a) Projected density of states (PDOS) of the Janus MoSSe/silk heterostructure showing contributions from the major atomic orbitals. (b) 2D planar charge density difference map displaying interfacial charge redistribution. The color bar ranges from 0 to 1, representing low to high charge density, respectively. Red/yellow regions indicate charge depletion, while blue/purple areas correspond to charge accumulation.}
	    \label{pdos-planar-charge}
\end{figure}

Figure \ref{pdos-planar-charge} presents the electronic structure and charge redistribution behavior of the Janus MoSSe/silk van der Waals heterostructure. Panel (a) shows the projected density of states (PDOS), where the valence band region (-4 to 0 eV) is dominated by the p-orbitals of C, N, O, S, and Se from both the silk polymer and the MoSSe monolayer. The upper portion of the valence band receives significant contributions from the S(p) and Se(p) states, while the deeper valence states arise mainly from the C(p) and N(p) orbitals of silk. In contrast, the conduction band manifold is governed primarily by Mo(d) orbitals, exhibiting well-defined peaks characteristic of the intrinsic MoSSe electronic structure. The absence of midgap states confirms that interface formation does not introduce defect levels or trap states, and the combined system maintains a type-II band alignment, with the silk layer contributing mainly to valence states and MoSSe defining the conduction edge. Panel (b) displays the 2D planar charge density difference distribution obtained by subtracting the isolated components from the combined heterostructure. The color scale spans from 0 (low charge density) to 1 (high charge density), enabling clear visualization of charge accumulation (blue/purple) and depletion (red/yellow). A distinct polarization pattern emerges at the interface, where electron accumulation is observed near the Mo and chalcogen atoms of the MoSSe layer, while the silk polymer undergoes local charge depletion around its electronegative N and O atoms. This asymmetric redistribution arises from the intrinsic out-of-plane dipole of Janus MoSSe synergistically interacting with the heterogeneous bonding landscape of the silk chain. The resulting interfacial dipole-dipole coupling enhances the electrostatic contrast at the junction, which is crucial for the triboelectric response and directly impacts charge transfer behavior under strain.

\begin{figure*}[ht]
\centering
\includegraphics[width=0.99\linewidth]{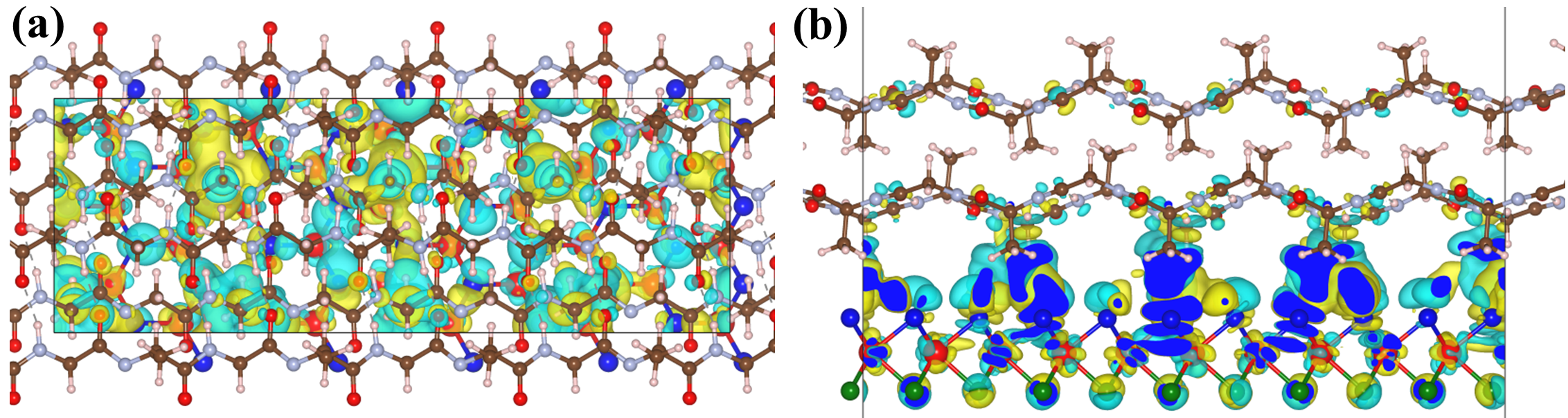}
\caption{3D charge density difference (CDD) distribution of the Janus MoSSe/silk heterostructure. Yellow and blue isosurfaces represent charge accumulation and depletion regions, respectively. The interfacial redistribution reveals that electrons transfer from the silk polymer to the Janus MoSSe surface. Bader charge analysis confirms a net electron transfer of 0.105854~$e$/atom from silk toward the MoSSe layer, indicating strong interfacial polarization driven by the intrinsic dipole of Janus MoSSe and the electronegative functional groups in silk.}
\label{cddp}
\end{figure*}

Figure \ref{cddp} illustrates the 3D charge density difference (CDD) of the Janus MoSSe/silk heterostructure, revealing the spatial distribution of charge accumulation and depletion at the interface. The yellow isosurfaces correspond to charge accumulation, while the blue regions represent charge depletion. This color-coded representation clearly captures how electrons and holes reorganize upon interface formation. A pronounced electronic asymmetry arises at the MoSSe/silk junction. The MoSSe surface-particularly around the Mo and chalcogen atoms-exhibits widespread yellow-colored accumulation zones. Conversely, the silk polymer shows blue depletion regions concentrated around its electronegative N and O atoms. This indicates that electrons migrate away from specific functional groups in the silk backbone and accumulate on the MoSSe side. To quantify this transfer, Bader charge analysis was performed. The results show that each silk atom donates approximately 0.11 e$^{-}$ to the Janus MoSSe surface. This confirms a net electron transfer from silk to MoSSe, consistent with the direction of the intrinsic dipole moment in Janus MoSSe. The charge redistribution enhances the built-in dipole at the interface, strengthening the electrostatic contrast between the two layers. The top view (see Figure \ref{cddp}a) highlights the delocalized nature of the accumulated charge spreading across the MoSSe surface, while the side view (see Figure \ref{cddp}b) provides a clear visualization of charge flow perpendicular to the interface. The layered pattern of positive and negative isosurfaces indicates strong dipole-dipole interaction between MoSSe and specific functional groups in the silk polymer. The 3D CDD analysis demonstrates that the MoSSe/silk interface undergoes substantial charge polarization, with electron flow from silk into the MoSSe layer. This interfacial polarization is expected to enhance triboelectric performance by increasing charge separation efficiency and modifying the local electrostatic environment under mechanical deformation.

\begin{figure}[htp!]
\centering
\includegraphics[width=1\linewidth]{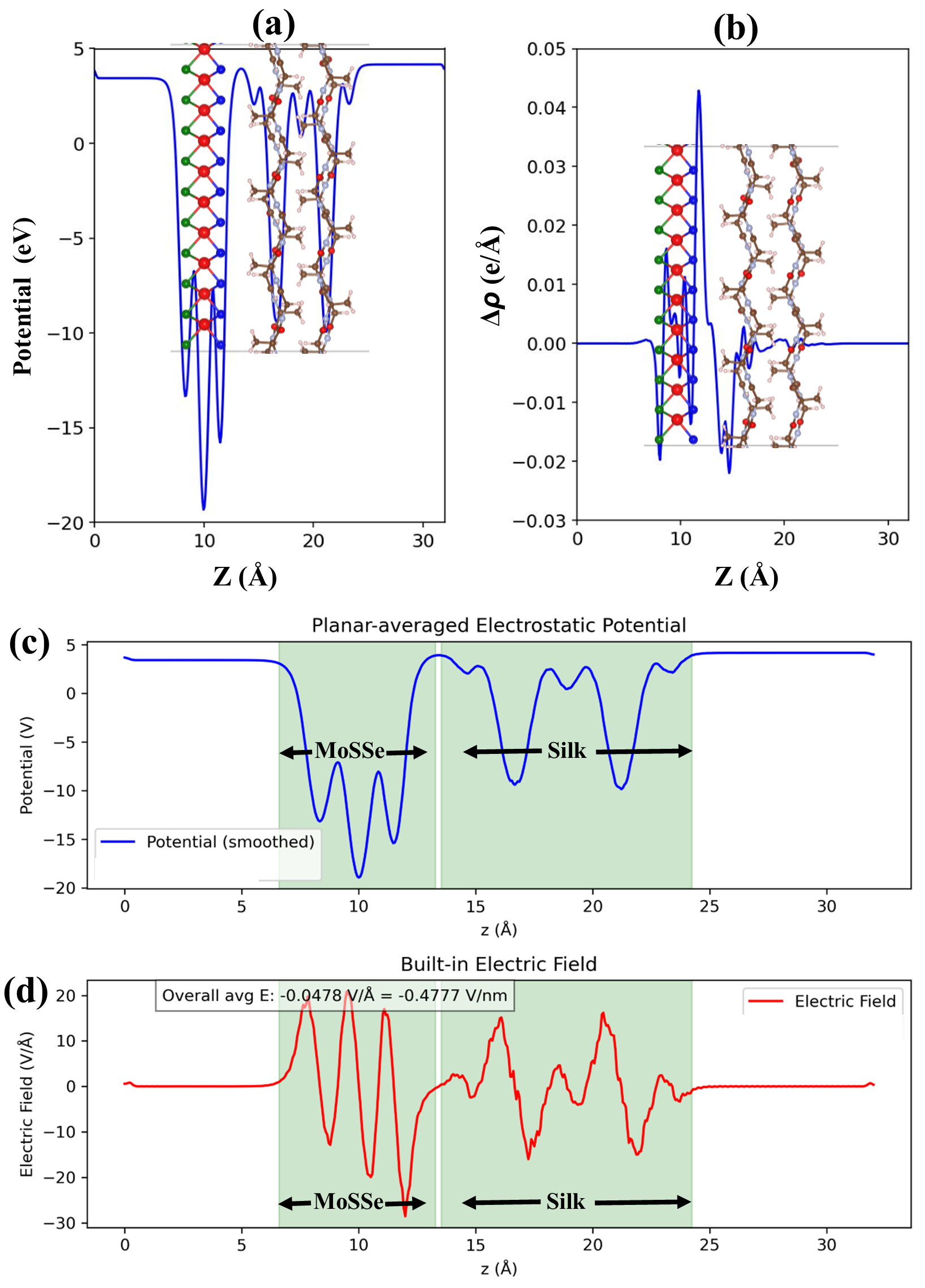}
\caption{Electrostatic characteristics of the Janus MoSSe/silk vdW heterostructure. (a) Planar-averaged electrostatic potential and (b) planar charge density difference illustrating the interfacial polarization. (c) Smoothed electrostatic potential profile across the MoSSe and silk regions. (d) Corresponding built-in electric field showing the internal field formed due to charge redistribution at the interface.}
\label{potential}
\end{figure} 

Figure \ref{potential} presents a detailed analysis of the electrostatic behavior of the Janus MoSSe/silk vdW heterostructure revealing how charge redistribution at the interface generates a built-in potential and internal electric field. Figure \ref{potential}(a) displays the planar-averaged electrostatic potential along the out-of-plane direction. The MoSSe layer exhibits pronounced potential wells associated with the asymmetric Mo-S and Mo-Se environments, while the silk polymer region shows a comparatively smoother potential landscape due to its organic backbone. The noticeable shift in potential between the two materials reflects the formation of an interface dipole once the heterostructure is formed.
Figure \ref{potential}(b) shows the planar charge density difference, highlighting the redistribution of electrons across the interface. The oscillatory patterns in $\Delta\rho$(z) indicate regions of charge accumulation and depletion, consistent with the characteristic behavior of Janus TMD interfaces. The silk polymer donates electrons toward the MoSSe surface, a behavior confirmed by Bader charge analysis, which indicates that each atom in silk transfers approximately 0.11 electrons to the MoSSe layer. This flow of electrons is driven by the electronegative functional groups within silk and the intrinsic vertical asymmetry of Janus MoSSe, together giving rise to significant interface polarization. Figure \ref{potential}(c) provides the smoothed electrostatic potential profile, demonstrating a continuous potential gradient across the heterostructure. The MoSSe region shows a deeper potential minimum, while the potential gradually rises through the silk region. This difference in vacuum levels enables extraction of work function values, the pristine Janus MoSSe monolayer shows a higher work function compared to pure silk, while the vdW heterostructure exhibits an intermediate value. The reduction in the heterostructure work function relative to pristine MoSSe arises directly from the electron donation from silk.  Figure \ref{potential}(d) illustrates the internal electric field derived from the potential gradient. A substantial built-in electric field develops at the interface due to charge transfer and the combined dipole moments of the two materials. The pristine Janus MoSSe monolayer exhibits a small but non-negligible dipole moment, while the silk polymer possesses an almost negligible intrinsic dipole. In contrast, the heterostructure shows a significantly enhanced dipole moment, reflecting strong interface-induced polarization. The corresponding internal electric field reaches nearly half a volt per nanometer, underlining the strength of the electrostatic interaction between the two layers. This analysis confirms that the MoSSe/silk heterostructure hosts strong interfacial polarization driven by electron transfer and dipole coupling, which is expected to play a major role in modulating charge separation and enhancing triboelectric nanogenerator performance. 

\begin{figure}[b]
\centering
\includegraphics[width=1\linewidth]{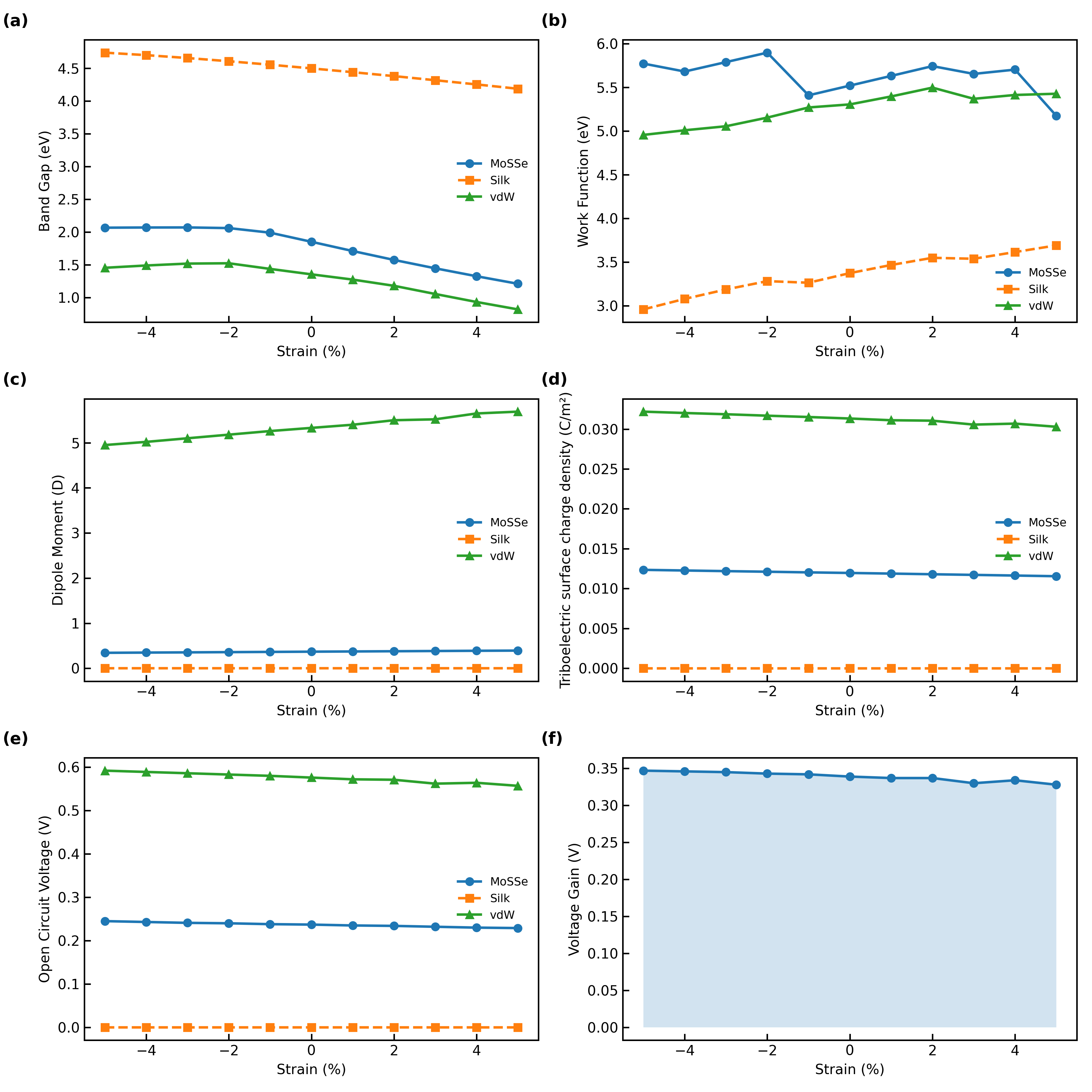}
\caption{Strain-dependent electronic and triboelectric properties of isolated Janus MoSSe, silk fibroin, and MoSSe/silk vdW heterostructure. (a) Band gap variation, showing significant reduction in the vdW heterostructure under tensile strain. (b) Work function evolution, indicating Fermi level alignment and interfacial charge transfer between MoSSe and silk. (c) Dipole moment (in Debye), revealing enhanced polarization in the vdW system due to asymmetric charge redistribution. (d) Triboelectric surface charge density, demonstrating a substantial increase in the vdW heterostructure compared to isolated constituents. (e) Open-circuit voltage ($V_{OC}$), showing improved output performance in the heterostructure. (f) Voltage gain ($\Delta V = V_{OC}^{vdW}-V_{OC}^{MoSSe}$), highlighting the enhancement in device performance induced by interfacial coupling and strain engineering.}
\label{TENGs}
\end{figure} 

\subsection{Strain-dependent triboelectric properties}

The strain-dependent electronic and triboelectric properties of Janus MoSSe, silk fibroin, and their vdW heterostructure are systematically analyzed to understand the role of interfacial interactions in enhancing device performance. As shown in Figure \ref{TENGs}(a), the band gap decreases with increasing tensile strain for all systems. Notably, the reduction is more significant in the MoSSe/silk vdW heterostructure compared to the isolated materials, indicating stronger interlayer electronic coupling under strain. While silk maintains a relatively large band gap due to its insulating nature, MoSSe shows moderate tunability, and the heterostructure exhibits enhanced sensitivity to strain, which facilitates charge transport.

The variation in work function (Figure \ref{TENGs}(b)) provides clear evidence of charge redistribution at the interface. The shift in work function observed in the vdW heterostructure, relative to the individual components, suggests the formation of an internal electric field due to Fermi level alignment. This built-in field drives electron transfer across the interface, leading to equilibrium charge redistribution, which is crucial for improving triboelectric performance. This interfacial charge transfer is further confirmed by the dipole moment results shown in Figure \ref{TENGs}(c). The vdW heterostructure exhibits a significantly higher dipole moment ($\sim$ 5~D) compared to isolated MoSSe ($\sim$0.35~D), while silk contributes negligibly. This large enhancement indicates strong polarization arising from asymmetric charge distribution at the interface. Additionally, a slight increase in dipole moment with tensile strain suggests that strain further enhances polarization effects.

The influence of polarization on charge accumulation is reflected in the triboelectric surface charge density (see Figure \ref{TENGs}(d)). The vdW heterostructure achieves values around $0.03~\mathrm{C/m^2}$, which are considerably higher than those of MoSSe ($\sim$0.012~$\mathrm{C/m^2}$) and silk (on the order of $10^{-7}~\mathrm{C/m^2}$). This substantial increase highlights the dominant role of interfacial coupling in enhancing charge storage capability. Furthermore, the gradual variation with strain indicates improved charge separation efficiency under mechanical deformation. Consequently, the open-circuit voltage ($V_{\mathrm{OC}}$) shown in Figure \ref{TENGs}(e) is significantly enhanced in the vdW heterostructure. While isolated MoSSe generates voltages in the range of $\sim$0.23-0.25~V and silk contributes negligibly, the heterostructure reaches values of $\sim$0.57-0.59~V. This improvement arises from the combined effects of increased dipole moment, enhanced surface charge density, and interfacial charge transfer.

Finally, the voltage gain ($\Delta V=V_{\mathrm{OC}}^{\mathrm{vdW}} - V_{\mathrm{OC}}^{\mathrm{MoSSe}}$) presented in Figure \ref{TENGs}(f) clearly demonstrates the performance enhancement achieved through vdW heterostructure formation. The consistently positive voltage gain across all strain values confirms that the vdW system outperforms the isolated MoSSe layer. This enhancement is primarily attributed to strain-modulated electronic properties and strong interfacial interactions. These results demonstrate that the MoSSe/silk vdW heterostructure exhibits superior triboelectric performance compared to its individual components. The combined effects of interfacial polarization, charge redistribution, and strain engineering make this system a promising candidate for next-generation triboelectric nanogenerators. 

\section{Conclusion}
\label{conclusion}
In summary, we have systematically investigated the Janus MoSSe/silk vdW heterostructure offers a strain-tunable pathway for achieving enhanced triboelectric nanogenerator performance. Tensile strain substantially modulates the electronic structure of the hetero-interface, leading to stronger band-gap reduction and more pronounced strain sensitivity than in the isolated MoSSe and silk systems. Interfacial Fermi-level alignment induces notable charge redistribution, which manifests as a significantly increased dipole moment and a marked enhancement in triboelectric surface charge density. These combined effects-interfacial polarization, strain-induced dipole reinforcement, and asymmetric charge transfer-collectively improve charge separation and elevate the open-circuit voltage in the heterostructure. The consistently positive voltage gain relative to pristine MoSSe confirms the superior performance of the vdW system across all strain conditions. This work highlights the importance of interfacial engineering in 2D/biopolymer hybrid systems and identifies the MoSSe/silk heterostructure as a promising candidate for flexible, sustainable, and high-efficiency triboelectric energy-harvesting devices.

\begin{acknowledgements} 
D.S. and R.L. thank the Carl Tryggers Stiftelse for Vetenskaplig Forskning (CTS 22:2283) for financial support. R.L. would like also to thank the {\AA}forsk Foundation (grant number 22-206). The computations were enabled by resources provided by the National Academic Infrastructure for Supercomputing in Sweden (NAISS), partially funded by the Swedish Research Council through grant agreement no. 2022-06725. NAISS (2023/22-1367 and 2024/5-38) Sweden is acknowledged for providing computing facilities. 
\end{acknowledgements} 
\bibliography{Ref}
\end{document}


\title{ {{\Huge{\underline{Supporting Information}} } \vspace*{0.8 cm}} \break Strain-tunable interface electrostatics in Janus MoSSe/silk vdW heterostructure for triboelectric nanogeneration}%

\author{Deobrat Singh}
\affiliation{KTH Royal Institute of Technology, Department of Materials Science and Engineering, Brinellvägen 23, SE-100 44 Stockholm, Sweden}
\email{deosing@kth.se}

\author{Raquel Liz\'arraga}
\affiliation{KTH Royal Institute of Technology, Department of Materials Science and Engineering, Brinellvägen 23, SE-100 44 Stockholm, Sweden}
\affiliation
{Wallenberg Initiative Materials Science for Sustainability, Department of Materials Science and Engineering, The Royal Institute of Technology, Stockholm SE-100 44, Sweden}

\date{\today}

	\date{\today}
	\maketitle
	\vspace*{1 cm}




\section{Electronic properties of Janus 2D MoSSe monolayer and silk type-II polymer}
The atom-projected band structure of the Janus 2D MoSSe monolayer shows a clear semiconducting nature with band edges mainly contributed by Mo d orbitals and S/Se p orbitals, reflecting its broken out-of-plane symmetry. The valence and conduction bands exhibit strong orbital hybridization and slight spin–orbit splitting near high-symmetry points. For the silk type-II system, the band structure indicates a wider distribution of states with dominant contributions from O, C, and N atoms, forming localized bands and a larger band gap characteristic of insulating organic materials.

\begin{figure}[htp!]
     \centering
\includegraphics[width=1.0\linewidth]{ 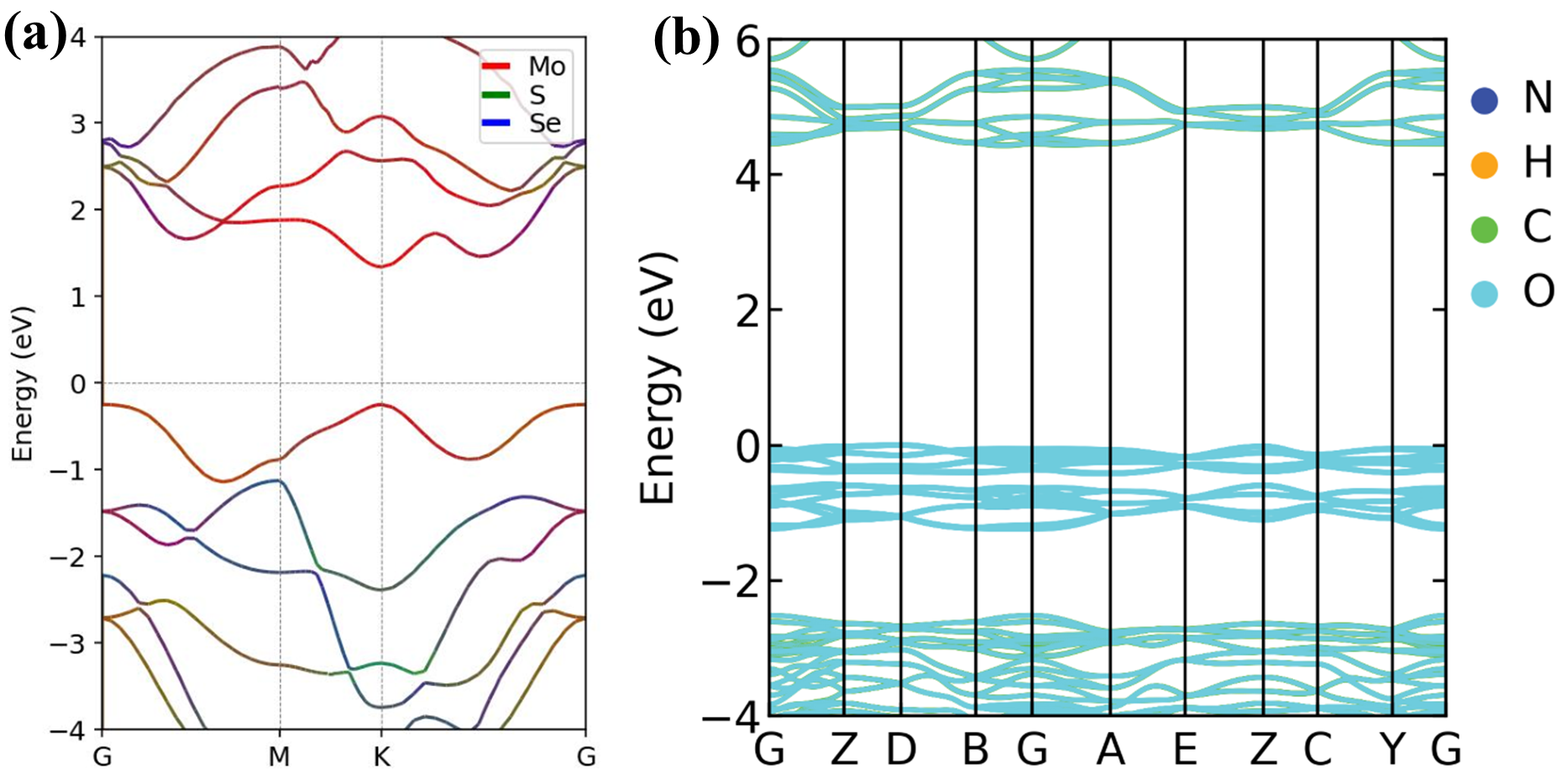}
\caption{Atom-resolved electronic band structures for pristine Janus 2D MoSSe monolayer (a) and pure silk type-II polymer structure (b).} 
\label{band}
\end{figure}
\clearpage

\newpage

    \begin{figure}[htp!]
	\centering
  \includegraphics[width=1.0\linewidth]{ 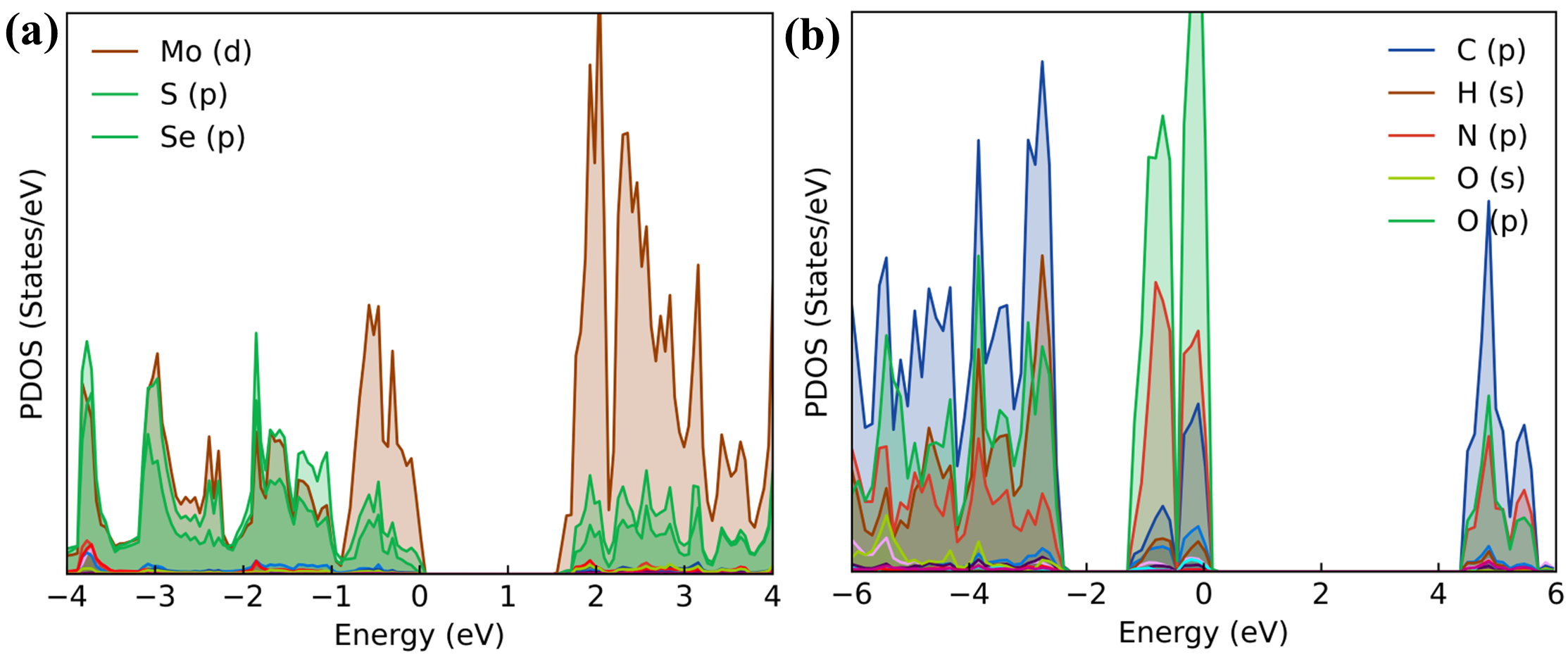}
	    \caption{Projected density of states (PDOS) for pristine Janus 2D MoSSe monolayer (a) and pure silk type-II polymer structure (b).} 
	    \label{pdos}
    \end{figure} 
\clearpage
\newpage

\section{{Ab-initio} molecular dynamics simulation of MoSSe/silk vdW heterostructure}
\begin{figure}[htp!]
	\centering
  \includegraphics[width=1.0\linewidth]{ 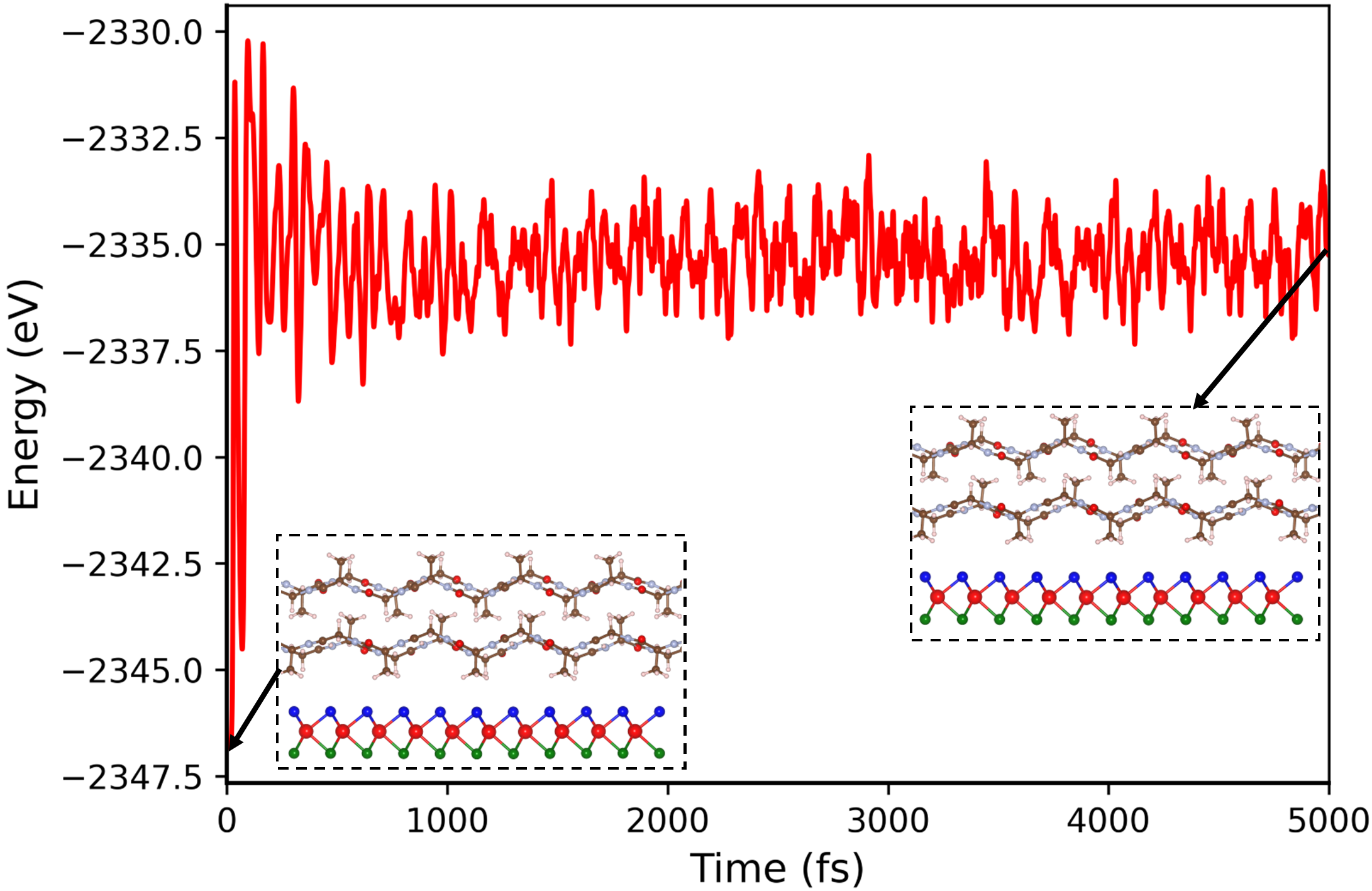}
	    \caption{Variation of total energy (eV) as a function of time (fs) obtained from \textit{ab-initio} molecular dynamics simulations of the of a MoSSe/silk vdW heterostructure, demonstrating the energetic stability of the system over the simulation time.} 
	    \label{aimd}
    \end{figure}

Figure \ref{aimd} shows the total energy as a function of simulation time for the MoSSe/silk vdW heterostructure at 300 K, computed via \textit{ab-initio} molecular dynamics using the NVT ensemble with a Nosé–Hoover chain thermostat. The energy exhibits only minor fluctuations around an equilibrium value and shows no systematic drift throughout the simulation, demonstrating good energy conservation and effective temperature control. This behavior confirms that the system is well equilibrated and that the vdW heterostructure maintains its structural integrity, highlighting the stability of the interfacial interactions governed by vdW forces.

